\documentclass{pasj01}
\SetRunningHead{J-GEM Observations for GW150914}
{Usage of \texttt{pasj00.cls}}

\begin{document}

\title{
J-GEM Follow-Up Observations 
to Search for an Optical Counterpart of 
The First Gravitational Wave Source GW150914
}
\author{Tomoki \textsc{Morokuma}\altaffilmark{1}}
\author{Masaomi \textsc{Tanaka}\altaffilmark{2}}
\author{Yuichiro \textsc{Asakura}\altaffilmark{3}}
\author{Fumio \textsc{Abe}\altaffilmark{3}}
\author{Paul J. \textsc{Tristram}\altaffilmark{4}}
\author{Yousuke \textsc{Utsumi}\altaffilmark{5}}
\author{Mamoru \textsc{Doi}\altaffilmark{1}}
\author{Kenta \textsc{Fujisawa}\altaffilmark{6}}
\author{Ryosuke \textsc{Itoh}\altaffilmark{7}}
\author{Yoichi \textsc{Itoh}\altaffilmark{8}}
\author{Koji S. \textsc{Kawabata}\altaffilmark{5}}
\author{Nobuyuki \textsc{Kawai}\altaffilmark{9}}
\author{Daisuke \textsc{Kuroda}\altaffilmark{10}}
\author{Kazuya \textsc{Matsubayashi}\altaffilmark{11}}
\author{Kentaro \textsc{Motohara}\altaffilmark{1}}
\author{Katsuhiro L. \textsc{Murata}\altaffilmark{12}}
\author{Takahiro \textsc{Nagayama}\altaffilmark{13}}
\author{Kouji \textsc{Ohta}\altaffilmark{11}}
\author{Yoshihiko \textsc{Saito}\altaffilmark{9}}
\author{Yoichi \textsc{Tamura}\altaffilmark{1}}
\author{Nozomu \textsc{Tominaga}\altaffilmark{14,15}}
\author{Makoto \textsc{Uemura}\altaffilmark{5}}
\author{Kenshi \textsc{Yanagisawa}\altaffilmark{10}}
\author{Yoichi \textsc{Yatsu}\altaffilmark{9}}
\author{Michitoshi \textsc{Yoshida}\altaffilmark{5}}
\altaffiltext{1}{Institute of Astronomy, Graduate School of Science, University of Tokyo, 2-21-1, Osawa, Mitaka, Tokyo 181-0015, Japan}
\altaffiltext{2}{National Astoronomical Observatory of Japan, Mitaka, Tokyo 181-8588, Japan }
\altaffiltext{3}{Institute for Space-Earth Environmental Research, Nagoya University, Chikusa-ku, Nagoya 464-8601, Japan}
\altaffiltext{4}{Mt. John University Observatory, Lake Tekapo 8770, New Zealand}
\altaffiltext{5}{Hiroshima Astrophysical Science Center, Hiroshima University, Hiroshima 739-8526, Japan}
\altaffiltext{6}{The Research Institute of Time Studies, Yamaguchi University, Yamaguchi 753-8511, Japan}
\altaffiltext{7}{Department of Physical Science, Hiroshima University, Hiroshima 739-8526, Japan}
\altaffiltext{8}{Nishi-Harima Astronomical Observatory, University of Hyogo, Hyogo 679-5313, Japan}
\altaffiltext{9}{Department of Physics, Tokyo Institute of Technology, Meguro-ku, Tokyo 152-8551, Japan}
\altaffiltext{10}{Okayama Astrophysical Observatory, National Astronomical Observatory of Japan, Asakuchi, Okayama 719-0232, Japan}
\altaffiltext{11}{Department of Astronomy, Kyoto University, Kitashirakawa-Oiwake, Kyoto 606-8502, Japan}
\altaffiltext{12}{Department of Particle and Astrophysical Science, Nagoya University, Chikusa-ku, Nagoya 464-8602, Japan}
\altaffiltext{13}{Graduate School of Science and Engineering, Kagoshima University, Kagoshima 890-0065, Japan}
\altaffiltext{14}{Department of Physics, Faculty of Science and Engineering, Konan University, Kobe, Hyogo 658-8501, Japan}
\altaffiltext{15}{Kavli Institute for the Physics and Mathematics of the Universe (WPI), The University of Tokyo, Kashiwa, Chiba 277-8583, Japan}
\email{tmorokuma@ioa.s.u-tokyo.ac.jp}


\maketitle

\begin{abstract}
We present our optical follow-up observations to search for an electromagnetic 
counterpart of the first gravitational wave 
source GW150914 
in the framework of the Japanese collaboration 
for Gravitational wave ElectroMagnetic follow-up (J-GEM), 
which is an observing group utilizing optical and radio telescopes 
in Japan, as well as those in New Zealand, China, South Africa, Chile, and Hawaii.
We carried out 
a wide-field imaging survey with Kiso Wide Field Camera (KWFC) on the 1.05-m Kiso Schmidt telescope in Japan 
and a galaxy-targeted survey with Tripole5 on the B\&C 61-cm telescope in New Zealand. 
Approximately 24 deg$^{2}$ regions in total 
were surveyed in $i$-band with KWFC and 
18 nearby galaxies 
were observed with 
Tripole5 in $g$-, $r$-, and $i$-bands 4-12 days after 
the gravitational wave detection. 
Median $5\sigma$ depths are 
$i\sim18.9$~mag for the KWFC data and 
$g\sim18.9$~mag, $r\sim18.7$~mag, and $i\sim18.3$~mag for the Tripole5 data. 
Probability for a counterpart to be in the observed area 
is 1.2\% in the initial skymap and 
0.1\% in the final skymap. 
We do not find any transient 
source associated to an external galaxy 
with spatial offset from its center, 
which is consistent with the local supernova rate. 
We summarize future prospects 
and ongoing efforts to pin down 
electromagnetic counterparts 
of binary black hole mergers 
as well as neutron star mergers. 
\end{abstract}

\section{Introduction}\label{sec:intro}

A new generation of 
gravitational-wave (GW) detectors, 
Advanced LIGO \citep{abbott2016a},
Advanced Virgo \citep{acernese2015},
and KAGRA \citep{somiya2012}, are designed to detect
GWs from mergers of neutron stars (NSs) and black holes (BHs).
These new detectors are 
much more sensitive than ever; 
with the design sensitivity,
the horizon distance will reach $\sim 200$~Mpc for NS-NS mergers 
and a few~Gpc for BH-NS or BH-BH mergers 
and 
many detections of GW events per year are expected \citep{abbott2016e}. 

Detections of electromagnetic (EM) counterparts are essential
to study their astrophysical properties and environments.
However, since positional localization only 
with the GW detectors is not accurate, which is 
larger than $100$ deg$^2$ 
during their early observing run (\cite{kasliwal2014}; \cite{singer2014})
and 
a few 10 deg$^2$ even in the LIGO-Virgo-KAGRA era 
\citep{nissanke2013,kelley2013}, 
it is a big challenge to identify the EM counterparts. 
To guide surveys for the EM identification, 
various kinds of EM signals have been theoretically
studied over a wide wavelength range,
from radio \citep{nakar2011},
infrared, optical, and ultraviolet 
\citep{li1998,kulkarni2005,metzger2010,tanaka2013}, 
X-ray \citep{nakamura2014,metzger2014magnetar}, 
and to gamma-ray (e.g., short gamma-ray burst; \cite{metzger2012}). 
Consequently, we have organized a group to carry out systematic follow-up observations of GW sources using Japanese facilities 
called ``Japanese collaboration for GW EM follow-up (J-GEM)'' as part of a larger worldwide EM follow-up effort.

Recently Advanced LIGO reported the first detection of the GW event 
(GW150914, \cite{abbott2016c}). 
The signal was detected 
at 09:50:45 on 2015 September 14 UT, 
4~days before the official start of the first observing run (O1) 
and an alert was delivered 
via GCN Notices in a machine-readable way at 03:12:12 and
by an e-mail manually at 05:39:44 on 2015 September 16 UT. 
The waveform indicates that the source of the GWs is a
merger of two BHs, whose masses are 
estimated to be $36^{+5}_{-4} M_{\odot}$
and $29^{+4}_{-4} M_{\odot}$, 
and that the luminosity distance is $410^{+160}_{-180}$ Mpc \citep{gw150914prop}. 
The position of the source is 
localized to 590 deg$^2$ (90 \% probability). 
We note that, ath the time of the initial alert,  
a classification of this GW source was not shared with the EM observers 
and the BH-BH nature was informed after our observations presented in this {\it Letter}. 

To search for an EM counterpart of GW150914,
extensive EM follow-up observations have been performed
following the alert \cite{abbott2016d}; \cite{abbott2016b}).
In this {\it Letter}, we report 
optical follow-up observations for GW150914 by the J-GEM collaboration. 
We refer to a skymap promptly produced with LALInference Burst (LIB; \cite{lynch2015}) 
as the initial skymap and to the most accurate LALInference \citep{veitch2015} 
skymap distributed on 2016 January 13 as the final skymap.
All the magnitudes shown in this {\it Letter} are in the AB system. 

\section{J-GEM Observations for GW150914}\label{sec:jgemobs}

J-GEM has observing facilities from radio to optical as listed 
in Table~\ref{tab:telall}. 
They are nicely distributed all over the Earth 
in terms of the longitude of the sites. 
Among them, we utilized two telescopes to carry out two types of optical follow-up observations 
for GW150914: 
one is an imaging survey 
with a wide-field imaging camera, 
Kiso Wide Field Camera (KWFC; \cite{sako2012}) 
mounted to the 1.05-m Kiso Schmidt telescope in Japan 
and the other is galaxy-targeted observations 
of nearby potential host galaxies of the GW source with 
Tripole5 on the 61-cm Boller \& Chivens (B\&C) Telescope 
at the Mt. John University Observatory in New Zealand. 
Most of the high probability regions are in the southern hemisphere 
and are difficult to observe with most of our observing facilities. 
Subaru Hyper Suprime-Cam (HSC; \cite{miyazaki2012}), 
which has the widest field-of-view among 8m-class telescopes, 
was not available after the alert until early October.

\subsection{Kiso KWFC Observations}\label{sec:jgemobs_survey}

KWFC is a wide-field optical imaging camera  
on the 1.05-m Kiso Schmidt telescope. 
The camera consists of eight 2k$\times$4k CCDs and 
the total field-of-view is 
$2.2$~deg~$\times\ 2.2$~deg. 

The KWFC observations were carried out on 2015 September 18, 
4.4~days after the GW detection. 
We took 
180-second
exposures for five continuous field-of-views, 
approximately $24$ deg$^{2}$ in total \citep{morokuma2016}. 
The observed area is shown in Figure~\ref{fig:obsfield} and 
details of the observations 
are summarized in Table~\ref{tab:obs_kiso}. 
High probability region in the skymap 
visible from the site during the night are almost towards the Sun and 
the target fields are observable only 
at very low elevation (high airmass: $\sec{(z)}>3$, where $z$ is the zenith distance) 
right before sunrise and during the astronomical twilight. 
Therefore, we chose the $i$-band filter to avoid high sky background due to the Sun as much as possible. 

The total probability of the regions observed with KWFC 
to include the GW source 
was initially 1.2\% but  
turned out to be $0.1$\% in the final skymap. 
The observed regions are partly overlapped with regions covered by 
Pan-STARRS (PS1; \cite{smartt2016}) 
and MASTER-NET \citep{lipunov2016}\footnote{http://master.sai.msu.ru/static/G184098/G184098\_4.png}. 

The data reduction procedure basically follows 
that of the supernova (SN) survey with KWFC (KISS; \cite{morokuma2014}). 
The $5\sigma$ limiting magnitudes 
are approximately 19~mag 
for the first four images and as shallow as 16.2~mag for the last image due to the twilight. 
For each of the fully reduced images, we applied an image subtraction method 
({\it hotpants}\footnote{http://www.astro.washington.edu/users/becker/v2.0/hotpants.html}) 
with deeper archival Sloan Digital Sky Survey (SDSS) images taken several years ago as references. 
Then, we extract transient objects with positive fluxes 
(2.5~$\sigma$, 5~pixel connection) 
in the subtracted images with {\it SExtractor} \citep{bertin1996}. 

\subsection{B\&C 61-cm Tripole5 Observations}\label{sec:obs_bc61cm}

Tripole5 is an optical camera on the B\&C 61cm telescope 
capable of taking images in 
$6.2$~arcmin~$\times~4.2$~arcmin field-of-view 
in $g$-, $r$-, and $i$-bands, simultaneously. 

\begin{figure}[!htbp]
 \begin{center}
  \includegraphics[width=81mm]{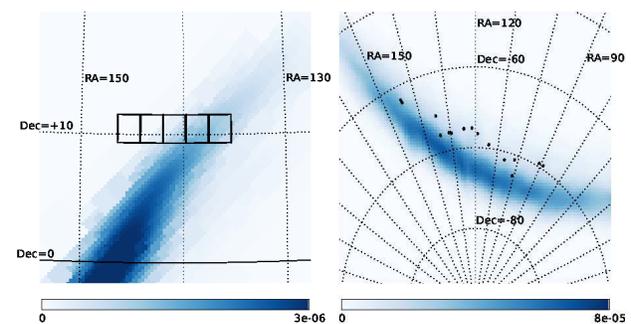}
 \end{center}
\caption{
Final skymap (LALInference) for the 
GW150914 localization 
and the observed regions with KWFC (left) and 
Tripole5 (right). 
The color map is shown in unit of probability per 
HEALPix \citep{gorski2005} pixel of $N_{\rm{side}}=2^{9}=512$, corresponding to about 47 arcmin$^{2}$. 
The KWFC field-of-views are shown in box and 
the area observed with Tripole5 are shown in dots. 
}
\label{fig:obsfield}
\end{figure}

The observations are started on 2015 September 20, 
6.3~days after the GW detection. 
We observed 18~nearby galaxies in the high probability region of the southern hemisphere 
as shown in Figure~\ref{fig:obsfield} and Table~\ref{tab:obs_bc61cm}. 
Two to six 
120-sec frames were taken per galaxy 
in $g$-, $r$-, and $i$-bands 
on 2015 September 20, 21, 24, and 26. 
The observed galaxies are selected from 
the Gravitational Wave Galaxy Catalogue 
(GWGC; \cite{white2011}) 
based on the initial skymap 
and are closer than 100~Mpc so that an EM counterpart of an NS-NS merger could be detected.
All the galaxies observed are located within the $\sim200$~deg$^{2}$ of the overlapped localization region (90\%~confidence) 
of GW150914 and 
GW150914-GBM, detected with 
Gamma-ray Burst Monitor (GBM) 
onboard the Fermi Gamma-ray Space Telescope \citep{connaughton2016}. 

The total probabilities in the initial and final skymaps are $0.003$\% 
although the distance $d=410^{+160}_{-180}$~Mpc is farther 
than the maximum distances to the galaxies by a factor of $\sim4$. 
The number of the observed galaxies is 
about 4\% of the galaxies in the GWGC catalog within the 90\% probability region.

The data are reduced in a standard manner using {\it IRAF}. 
Zeropoint magnitudes 
in the $g, r,$ and $i$-bands 
are determined relative to the $B$, $R$, and $I$-band magnitudes of objects in the USNO-B1.0 catalog \citep{monet2003} 
using the conversion equations in \citet{fukugita1996}. 
Medians of the $5\sigma$ limiting magnitudes are 
$g=18.9$ mag, 
$r=18.7$ mag, 
and $i=18.3$ mag.
Object catalogs for the Tripole5 images are created using {\it SExtractor}.

\section{Results \& Discussion}\label{sec:results_discussion}

For the KWFC data, 
radial profiles and SDSS classifications (star/galaxy separation based on {\it probPSF} information available in the SDSS database) of 
all of the transient objects 
are used to extract extragalactic transients. 
Known asteroids are also checked with {\it MPChecker} and removed from the transient catalog. 
For the Tripole5 data, 
the object catalog in the entire observed field of each target galaxy is first compared with the USNO-B1.0 catalog. 
There remain some objects without any counterparts in the USNO-B1.0 catalog 
and they are visually inspected by comparing the Tripole5 images with the Digitized Sky Survey images. 

In these procedures described above, 
we find no extragalactic transient object 
with a spatial offset from its host galaxy although we detect 
variability at centers of several external galaxies (including PS15cek described below).
Given the survey areas, depths of the data, and the measured volumetric 
SN rates in the local universe 
(\cite{blanc2004} for Type~Ia SNe and \cite{li2011} for Core-Collapse SNe), 
expected number of SNe is smaller than unity. 
This is consistent with the result that we do not find any SN candidate. 

Among transient objects discovered and reported by other projects so far 
(PS1 by \cite{smartt2016}; 
UVOT on the Swift satellite by 
\cite{evans2016}; 
La Silla - QUEST by \cite{rabinowitz2015}, 
and iPTF by \cite{singer2015} and \cite{kasliwal2016})
four transients are within 
the regions observed with KWFC; 
PS15cej, 
PS15cek, 
PS15ckf, and 
PS15dft (Table~\ref{tab:transients}). 
All these PS1 transients were discovered in early October 2015, 
while our KWFC data were 
taken in September 2015. 
PS15cek is a known AGN (2MFGC~07447; \cite{veroncetty2001}) at $z=0.060$, 
and the variability is also detected in our KWFC images compared with the past SDSS data.
We do not detect the other three objects, i.e.,
two SNe (PS15cej and PS15ckf) and one cataclysmic variable star
(PS15dft or ASASSN-15se), but the non-detection of variability is not surprising
since these explosions and flare are likely to occur after
our observations in September \citep{smartt2016}.

Several theoretical scenarios 
about EM counterparts of BH-BH mergers and their emissions are available. 
\citet{nakamura2016} calculated an expected optical emission from 
almost the same system as GW150914. 
In their model, Eddington luminosity of 
a $60$~M$_\odot$ BH in dense interstellar medium 
may have brightness of about 26~mag 
if the emission is radiated mainly in the optical wavelength at a similar distance to GW150914 ($d\sim300$~Mpc), 
which can be detected with 
8m-class telescopes and instruments. 
\citet{yamazaki2016} and \citet{morsony2016} predict 
a wide range of EM emissions from GW150914 
under an assumption that GW150914-GBM \citep{connaughton2016} is associated with the GW150914 \citep{loeb2016}.
They suggest that 
an optical counterpart is detectable with 8-m class telescopes 
within 1 or a few days 
after the GW event 
and that earlier emission within several hours 
after the event can be detected 
with smaller aperture (2-4~m) telescopes. 

These theoretical models indicate that wide-field surveys with 
8m-class telescopes and instruments such as Subaru HSC 
or Large Synoptic Survey Telescope (LSST; \cite{lsst2009}) 
is the best strategy 
to detect an optical counterpart of a GW event like GW150914. 
If a similar event to GW150914 occurs at a closer distance, 
an optical counterpart could be detectable with 1-2-m class telescopes. 
Although selecting the counterpart from a huge amount of imaging data is challenging, 
several intensive works have been done. 
Implementation of machine-learning 
techniques for reducing real-to-bogus ratios
with wide-field imaging data 
have been done in several projects (e.g., \cite{bloom2012}; \cite{brink2013}). 
Among the J-GEM instruments, 
machine-learning approach for effective discoveries of transient objects is being done 
for Subaru HSC data \citep{morii2016}.
Effective classification trial for many transient objects \citep{kessler2010} 
by adding realistic theoretical models for GW EM counterparts including NS-NS merger 
(see the review by \cite{tanaka2016})
will also help us to identify the EM counterpart in near future. 

\begin{ack}
This work is supported by MEXT Grant-in-Aid for Scientific Research on Innovative Areas 
``New Developments in Astrophysics Through Multi-Messenger Observations of Gravitational Wave Sources'' 
(24103003) 
and its Koubo Researches (25103503, 15H00788, 15H00774). 
This work is also supported by JSPS (15H02075). 
This study utilizes the archival images 
from the Sloan Digital Sky Survey 
and those from the Digitized Sky Surveys. 
Full acknowledgments can be found at 
http://www.sdss.org/collaboration/credits.html and 
https://archive.stsci.edu/dss/acknowledging.html, respectively. 

\end{ack}





\begin{table*}[!htbp]
  \tbl{J-GEM Telescopes in an order of longitude.}{%
  \begin{tabular}{lccccccc}  
\hline\noalign{\vskip3pt} 
 \multicolumn{1}{c}{Site (telescope)} & $d$ [m]$^{a}$ & Location$^{b}$ & Instrument$^{*}$ & FoV & Pixel Scale$^{c}$ & Note$^{f}$ \\   [2pt] 
\hline\noalign{\vskip3pt} 
Mt. Johns      	(B\&C 61cm) 	& 0.61        	& 170.47~E, 43.40~S, 1029 	& Tripole5      	& \timeform{4.'2} $\times$ \timeform{6.'2} 	& 0.17 	& (1)	\\ 
Mt. Johns      	(MOA-II) 	& 1.8         	& 170.47~E, 43.40~S, 1029 	& MOA-cam3 [1] 	& \timeform{1.31D} $\times$ \timeform{1.64D} 	& 0.58 	& (3)	\\ 
Akeno          	(MITSuME) 	& 0.5         	& 138.48~E, 35.79~N, 900  	& ($g,R_C,I_C$ imager) 	& \timeform{27.8'} $\times$ \timeform{27.8'} 	& 1.63     	& (1)	\\ 
Kiso           	(Kiso Schimidt) 	& 1.05        	& 137.63~E, 35.79~N, 1130 	& KWFC [2] 	& \timeform{2.2D} $\times$ \timeform{2.2D}      	& 0.946 	& (3)	\\ 
Nishi-Harima   	(Nayuta) 	& 2.0         	& 134.34~E, 35.03~N,  449 	& MINT        	& \timeform{10.9'} $\times$ \timeform{10.9'} 	& 0.32  	& (1)	\\ 
Okayama, OAO   	(Kyoto-3.8m$^{d}$) 	& 3.8      	& 133.60~E, 34.58~N,  343 	& KOOLS-IFU   	& \timeform{14''} $\phi$  	& 1.14 	& (2)	\\ 
Okayama, OAO   	(OAO~188cm) 	& 1.88        	& 133.59~E, 34.58~N,  371 	& KOOLS-IFU   	& \timeform{30''} $\phi$  	& 2.34 	& (2)	\\ 
Okayama, OAO   	(OAO~91cm) 	& 0.9         	& 133.59~E, 34.58~N,  364 	& OAO-WFC [3] 	& \timeform{28.4'} $\times$ \timeform{28.4'} 	& 1.67 	& (1)	\\ 
Okayama, OAO   	(MITSuME) 	& 0.5         	& 133.59~E, 34.58~N,  358 	& ($g,R_C,I_C$ imager) [4],[5] 	& \timeform{26.9'} $\times$ \timeform{26.9'} 	& 1.52 	& (1)	\\ 
Higashi-Hiroshima 	(Kanata) 	& 1.5         	& 132.78~E, 34.38~N,  511 	& HOWPol [6]    	& \timeform{15'} $\phi$               	& 0.30       	& (1)	\\ 
Higashi-Hiroshima 	(Kanata) 	& 1.5         	& 132.78~E, 34.38~N,  511 	& HONIR [7],[8] 	& \timeform{10'} $\times$ \timeform{ 10'}       	& 0.30       	& (1)	\\ 
Yamaguchi    	(Yamaguchi$^{e}$) 	& $32\times2$ 	& 131.56~E, 34.22~N,  166 	& 6-8~GHz Receiver     	& -                         	& 4-5 arcmin 	& (1)	\\ 
Tibet          	(HinOTORI$^{d}$) 	& 0.5         	&  80.03~E, 32.31~N, 5130 	& ($u,R_C,I_C$ imager) 	& \timeform{24'} $\times$ \timeform{24'} 	& 0.68       	& (1)	\\ 
Sutherland, SAAO 	(IRSF) 	& 1.4         	&  20.81~E, 32.38~S, 1761 	& SIRIUS [9],[10] 	& \timeform{7.7'} $\times$ \timeform{7.7'}    	& 0.45       	& (1)	\\ 
Pampa la Bola  	(ASTE$^{e}$) 	& 10          	&  67.70~W, 22.97~S, 4862 	& ASTECAM [11]    	& \timeform{8.1'}~$\phi$            	& 20-30      	& (1)	\\ 
Chajnantor, TAO	(miniTAO) 	& 1.04        	&  67.74~W, 22.99~S, 5640 	& ANIR [12]       	& \timeform{5.1'} $\times$ \timeform{ 5.1'}   	& 0.298      	& (1)	\\ 
Mauna Kea, MKO 	(Subaru) 	& 8.2         	& 155.48~W, 19.83~N, 4139 	& HSC [13]        	& \timeform{1.5D} $\phi$   	& 0.168      	& (3)	\\    [2pt] 
\hline 
    \end{tabular}}\label{tab:telall}
\begin{tabnote}
References for instruments; 
[1]: \citet{sako2008}, 
[2]: \cite{sako2012}, 
[3]: \citet{yanagisawa2014}, 
[4]: \citet{kotani2005}, 
[5]: \citet{yanagisawa2010}, 
[6]: \citet{kawabata2008}, 
[7]: \citet{akitaya2014}, 
[8]: \citet{sakimoto2012}, 
[9]: \citet{nagashima1999}, 
[10]: \citet{nagayama2003}, 
[11]: \citet{oshima2012}, 
[12]: \citet{konishi2015}, 
[13]: \citet{miyazaki2012}. \\
a: diameter of telescope primary mirror. \\
b: longitude and latitude in degrees, and height in meters. \\
c: pixel scale is in arcsec pixel$^{-1}$. \\
d: to be operated. \\
e: radio or submillimeter telescopes. \\
f: 
(1) galaxy-targeted;  
(2) integral field spectroscopy;  
(3) wide-field survey.   
\end{tabnote}   
\end{table*}

\begin{table*}[!htbp]
  \tbl{Summary of Kiso KWFC Observations. }{%
  \begin{tabular}{cccccccc}
      \hline
      UT$^{a}$ & MJD$^{b}$ & Field & RA & Dec & Filter & $t_{\rm{exp}}$$^{c}$ & $m_{\rm{lim}}(5\sigma)$\\
      \hline
	2015-09-18 19:06:02 & 57283.7969 & KT009891 & 09:05:35.52 & +10:27:00.0 & $i$ & 180 & 19.2\\
	2015-09-18 19:11:37 & 57283.8008 & KT009892 & 09:14:32.16 & +10:27:00.0 & $i$ & 180 & 18.9\\ 
	2015-09-18 19:16:59 & 57283.8045 & KT009893 & 09:23:28.80 & +10:27:00.0 & $i$ & 180 & 18.9\\
	2015-09-18 19:22:22 & 57283.8083 & KT009894 & 09:32:25.44 & +10:27:00.0 & $i$ & 180 & 18.9\\
	2015-09-18 19:34:29 & 57283.8167 & KT009895 & 09:41:22.08 & +10:27:00.0 & $i$ & 180 & 16.3\\
      \hline
    \end{tabular}}\label{tab:obs_kiso}
\begin{tabnote}
a: time starting the exposures. 
b: MJDs of the middle of the exposures. 
c: Exposure time for each field is in seconds. 
\end{tabnote}
\end{table*}

\begin{table*}[!htbp]
  \tbl{Summary of B\&C 61cm telescope Tripole5 Observations. }{%
  \begin{tabular}{ccccccccccc}
      \hline
      Galaxy & RA$^{a}$ & Dec$^{a}$ & $d$~[Mpc]$^{a}$ & Filter & 2015-09-20$^{b}$ & 2015-09-21$^{b}$ & 2015-09-24$^{b}$ & 2015-09-26$^{b}$\\
      \hline
      ESO034-012 & 06:43:30.8 & $-$72:35:41 & $74.22\pm11.13$ & $g,r,i$ & - & $120\times2$ & $120\times4$ & $120\times4$\\
      ESO058-014 & 06:46:36.1 & $-$70:36:54 & $93.42\pm14.01$ & $g,r,i$ & - & $120\times2$ & $120\times4$ & $120\times4$\\
      ESO058-023 & 07:04:45.5 & $-$71:00:59 & $92.83\pm13.93$ & $g,r,i$ & - & $120\times2$ & $120\times4$ & $120\times4$\\
      ESO059-023 & 07:56:06.1 & $-$68:16:41 & $70.51\pm10.58$ & $g,r,i$ & - & $120\times2$ & $120\times4$ & $120\times4$\\
      ESO060-010 & 08:38:36.7 & $-$67:56:11 & $96.96\pm14.54$ & $g,r,i$ & - & $120\times2$ & $120\times4$ & $120\times4$\\
      ESO060-011 & 08:42:43.0 & $-$67:48:54 & $94.97\pm14.25$ & $g,r,i$ & $120\times2$ & $120\times2$ & $120\times4$ & $120\times4$\\
      ESO060-018 & 08:56:40.5 & $-$67:52:13 & $84.89\pm12.73$ & $g,r,i$ & - & $120\times2$ & $120\times4$ & $120\times4$\\
      ESO089-009 & 08:05:09.0 & $-$67:35:12 & $95.29\pm20.96$ & $g,r,i$ & $120\times4$ & $120\times2$ & $120\times6$ & $120\times4$\\
      ESO089-015 & 08:18:08.1 & $-$67:34:37 & $96.53\pm21.24$ & $g,r,i$ & - & $120\times2$ & $120\times4$ & $120\times4$\\
      ESO089-016 & 08:18:23.4 & $-$67:36:40 & $97.78\pm21.51$ & $g,r,i$ & $120\times2$ & $120\times2$ & $120\times4$ & $120\times4$\\
      ESO090-011 & 08:58:18.5 & $-$65:22:03 & $72.93\pm10.94$ & $g,r,i$ & $120\times2$ & $120\times2$ & $120\times4$ & $120\times4$\\
      ESO126-023 & 09:37:51.2 & $-$62:09:04 & $33.42\pm5.01$  & $g,r,i$ & - & $120\times2$ & $120\times4$ & $120\times4$\\
      ESO126-024 & 09:38:29.1 & $-$61:49:47 & $33.42\pm5.01$  & $g,r,i$ & $120\times2$ & $120\times2$ & $120\times4$ & $120\times4$\\
      NGC~2150   & 05:55:46.3 & $-$69:33:39 & $58.89\pm8.83$  & $g,r,i$ & $120\times3$ & $120\times2$ & $120\times4$ & $120\times4$\\
      NGC~2187   & 06:03:48.5 & $-$69:35:00 & $59.56\pm8.93$  & $g,r,i$ & $120\times4$ & $120\times2$ & $120\times4$ & $120\times4$\\
      NGC~2187A  & 06:03:44.2 & $-$69:35:18 & $50.93\pm11.21$ & $g,r,i$ & $120\times2$ & $120\times2$ & $120\times4$ & $120\times4$\\
      NGC~2442   & 07:36:23.8 & $-$69:31:51 & $17.30\pm2.60$  & $g,r,i$ &  -           & $120\times3$ & $120\times4$ & $120\times4$\\
      NGC~2466   & 07:45:16.0 & $-$71:24:38 & $70.86\pm10.63$ & $g,r,i$ & $120\times2$ & $120\times2$ & $120\times4$ & $120\times4$\\
      \hline
    \end{tabular}}\label{tab:obs_bc61cm}
\begin{tabnote}
a: The coordinates of the center of the galaxy and the distances $d$ to the galaxies are derived from the GWGC \citep{white2011}. \\
b: Exposure time on each date for each galaxy is in seconds. 
\end{tabnote}
\end{table*}

\begin{table*}[!htbp]
  \tbl{Transients Reported in Other Papers in Our Survey Fields.}{%
  \begin{tabular}{cccccccccc}
      \hline
      Name & RA & Dec & Reference & Nature & Disc$^{a}$ & ID$^{b}$ & Expl.$^{c}$ & J-GEM & note\\
      \hline
		PS15cej & 09:35:19.41 & $+$10:11:50.7 & \citet{smartt2016} & SN~Ia & 2015-10-02 & 2015-10-10 & 2015-09-22 & KWFC & before Expl.\\
		PS15cek & 09:36:41.04 & $+$10:14:16.2 & \citet{smartt2016} & AGN   & 2015-10-02 & - & - & KWFC & -\\
		PS15ckf & 09:45:57.71 & $+$09:58:31.4 & \citet{smartt2016} & SN~II & 2015-10-03 & 2015-10-20 & 2015-09-27 & KWFC & before Expl.\\
		PS15dft & 09:33:09.38 & $+$10:28:02.2 & \citet{smartt2016} & CV    & 2015-10-23 & - & - & KWFC & -\\
      \hline
    \end{tabular}}
    \label{tab:transients}
\begin{tabnote}
a: PS1 discovery date. \\
b: Date of PS1 spectroscopic identification. \\
c: Explosion date based on the spectroscopic phase. 
\end{tabnote}
\end{table*}

\end{document}